\documentclass[twocolumn]{aastex62}
\usepackage{amsmath,graphicx,epsfig,apjfonts}

\newcommand{\PMO}{Purple Mountain Observatory, Chinese Academy of Sciences, Nanjing 210008, China}
\newcommand{\USTC}{School of Astronomy and Space Science, University of Science and Technology of China, Hefei, Anhui 230026, China}

\begin{document}

\title{The Energy Function and Cosmic Formation Rate of Fast Radio Bursts}

\author{Can-Min Deng}
\altaffiliation{dengcm@pmo.ac.cn}
\affiliation{\PMO}
\affiliation{\USTC}

\author{Jun-Jie Wei}
\altaffiliation{jjwei@pmo.ac.cn}
\affiliation{\PMO}

\author{Xue-Feng Wu}
\altaffiliation{xfwu@pmo.ac.cn}
\affiliation{\PMO}
\affiliation{\USTC}

\begin{abstract}
Fast radio bursts (FRBs) are intense radio transients whose physical origin remains unknown.
Therefore, it is of crucial importance to use a model-independent method to obtain the energy function
and cosmic formation rate directly from the observational data. Based on current samples from the Parkes and ASKAP telecsopes,
we determine, for the first time, the energy function and formation rate of FRBs by using the Lynden-Bell $\rm C^{-}$ method.
The energy function derived from the Parkes sample is a broken power law, however it is a simple power law for the ASKAP sample. For Parkes sample,
we derive the formation rate which is roughly consistent with the star formation rate up to $z\sim1.7$, with a local formation rate of $\dot \rho (0) \simeq (3.2\pm 0.3)\; \times {10^4}\;{\rm{Gp}}{{\rm{c}}^{ - 3}}{\rm{y}}{{\rm{r}}^{ - 1}}$ above a detection threshold of $2\,{\rm{Jyms}}$. For ASKAP sample, we find that the formation rate evolves much faster than the star formation rate up to $z\sim0.7$, namely $\dot \rho (z) \propto {(1 + z)^{6.9 \pm 1.9}}$, with a local formation rate of $\dot \rho (0) \simeq (4.6\pm 0.8)\; \times {10^3}\;{\rm{Gp}}{{\rm{c}}^{ - 3}}{\rm{y}}{{\rm{r}}^{ - 1}}$ above a detection threshold of $51\,{\rm{Jyms}}$. This might be a important clue for the physical origin of FRBs.
\end{abstract}

\keywords{Fast radio burst, Formation rate, Energy function}

\section{Introduction}
\label{sect:intro}
Fast radio bursts (FRBs) are intense radio transients with intrinsic durations less than several milliseconds \citep{Lorimer2007, Thornton2013, Ravi2019}.
So far, more than seventy FRBs have been detected by various telescopes \citep{Petroff2016}. They are all non-repeating sources except FRB 121102 \citep{Spitler2016}
and FRB 180814 \citep{The CHIME/FRB Collaboration2019}.
Their dispersion measures (DMs) are typically hundreds of ${\rm{pc}}\;{\rm{c}}{{\rm{m}}^{{\rm{ - }}3}}$, which are much higher than the DM contribution of our Galaxy,
robustly suggest that FRBs occur at cosmological distances. This is supported by the identification of the host galaxy of the repeating FRB 121102 with redshift $z\simeq0.19$ \citep{Chatterjee2017,Tendulkar2017}.

Many models have been proposed to explain FRBs.
It was suggested that FRBs can be produced by emission from a single neutron star (NS),
such as giant pulses from young pulsars \citep{Keane2012, Cordes2016}, giant flares from magnetars \citep{Lyubarsky2014}, and
magnetic field shedding of collapsing neutron stars \citep{Zhang2014,Falcke2014,Punsly2016}.
Some compact binary mergers, specifically, mergers of double white dwarfs (WDs) \citep{Kashiyama2013},
of WD-NS \citep{Gu2016,Liu2018}, of WD--black hole (BH) \citep{Li2018}, of double NSs \citep{Totani2013, Wang2016}, NS--BH \citep{Mingarelli2015}, or of double BHs \citep{Liu2016, Zhang2016}, could be responsible for FRBs. Besides these, there are also other novel models, such as pulsar traveling though asteroid belt \citep{Dai2016}, super-conducting strings \citep{Cai2012}, primordial black holes coalescence \citep{Deng2018}, and so on \citep[see][for a comprehensive review]{Platts2018}. However, the nature of FRBs remains unknown. The formation rate and energy function as well as their cosmological evolution are crucial for revealing the origin of FRBs. In this paper, we investigate the formation rate and energy function of FRBs. The errors quoted through this paper are all at the 1$\sigma$ confidence level.

\section{SAMPLE SELECTION}
\label{sec:data}

To date, a total of more than seventy FRBs have been detected by various radio telescopes including the Arecibo, ASKAP,
CHIME, GBT, Parkes, Pushchino, and UTMOST, which are available from the FRB catalogue\footnote{FRB catalogue website http://frbcat.org/, the parameters of the corresponding telescopes can also be found in the website. }
(see \citealt{ Petroff2016} and references therein). Since different telescopes have different detection thresholds, here we only take into account the two subsamples out of the 28 FRBs detected by the Parkes and 25 detected by the ASKAP telescopes respectively, to ensure that the very different response functions of different telescopes will not be involved.

The observed DM of an FRB can be consisted of
\begin{equation}
\rm DM_{obs}=DM_{MW}+DM_{IGM}+\frac{DM_{host}}{1+\emph{z}}\;,
\end{equation}
where
$\rm DM_{MW}$ is the DM distribution from the Milky Way and $\rm DM_{host}$ is the DM contribution
from both the FRB host galaxy and source environment in the cosmological rest frame of the FRB.
The IGM portion of DM is related to the redshift of the source through
\citep{Ioka2003, Inoue2004, Deng2014}
\begin{equation}
{\rm DM_{IGM}}(z)=\frac{3cH_0\Omega_b f_{\rm IGM} f_{e}}{8\pi G m_{p}}\int_{0}^{z}\frac{H_0(1+z')}{H(z')}dz'\;,
\end{equation}
where $\Omega_b$ is baryon density, $f_{\rm IGM}\sim0.83$ is the fraction of baryons in the IGM \citep{Fukugita1998},
and $f_{e}\sim7/8$ is the free electron number per baryon in the universe, which was first introduced
by \cite{Deng2014}. Adopting the latest Planck results for the $\Lambda$CDM cosmological parameters,
i.e., $H_0=67.74$ km $\rm s^{-1}$ $\rm Mpc^{-1}$, $\Omega_b=0.0486$, $\Omega_{\rm m}=0.3089$, and
$\Omega_{\Lambda}=0.6911$ \citep{Planck Collaboration2016}, the redshifts of the FRBs can then
be inferred from their $\rm DM_{IGM}$.
In order to obtain $\rm DM_{IGM}$ of an FRB, we have to figure out $\rm DM_{MW}$ and $\rm DM_{host}$.
The $\rm DM_{MW}$ values have been derived based on the Galactic electron density models of \cite{ Cordes2002}
or \cite{Yao2017}, as provided in the FRB catalogue (see \citealt{Petroff2016} and references therein),
while the $\rm DM_{host}$ values are mostly unknown. The observations of the host galaxy of FRB 121102
suggest that $\rm DM_{host}$ ($\sim 100$ pc $\rm cm^{-3}$) is not small, which is comparable to $\rm DM_{IGM}$
for FRB 121102 \citep{Tendulkar2017}. However, the surrounding environments of non-repeating FRBs
seem different with that of FRB 121102. Polarization measurements suggested that FRB 121102 is associated with an extreme
magneto-ionic environment \citep{Michilli2018}, but the non-repeaters are not \citep{Caleb2018}.
Furthermore, the search for the host galaxy of FRB 171020, possibly the most nearby FRB so far, suggested it might
be hosted in a low star-forming Sc galaxy and might not be associated with a luminous and compact radio continuum source \citep{Mahony2018},
in contrast to the case of FRB 121102. And we note that $\rm DM_{FRB\;171020}=DM_{obs}-DM_{MW}=76.1$ pc $\rm cm^{-3}$, which suggests the $\rm DM_{host}$ value for non-repeaters might be not large, again in contrast to the case of FRB 121102. Therefore, we adopt a fixed $\rm DM_{host}=50$ pc $\rm cm^{-3}$, which is close to $\rm DM_{MW}$, to derive a rough estimation on $\rm DM_{IGM}$, and hence, a rough estimation on $z$, of a particular FRB.

With an inferred redshift $z$, we calculate the isotropic energy\footnote{In this paper, we care more about the observed energy rather than about the observed luminosity.
That is because we note that the intrinsic pulse widths of most FRBs are unknown \citep{Ravi2019} and the observed energy is more physical than the observed luminosity. }
of an FRB within the rest-frame bandwidth $({\nu _{1 }},{\nu _{2}})$ by
\begin{equation}
E \simeq \frac{{4\pi D_L^2}}{{1 + z}}{F_v}\int_{{\nu _1}/(1 + z)}^{{\nu _2}/(1 + z)} {{{\left( {\frac{\nu }{{{\nu _{\rm{c}}}}}} \right)}^{ - \alpha }}d\nu }  ,
\label{eq:Eiso}
\end{equation}
where ${{F_\nu }}$ is the observed fluence density, ${{D_L}}$ is luminosity distance, ${{{\left( {\frac{\nu }{{{\nu _{\rm{c}}}}}} \right)}^{ - \alpha }}}$ is the spectrum of FRBs, $\nu_{c}=1.352$ GHz and $\nu_{c}=1.297$ GHz are the typical central frequency of Parkes and ASKAP, respectively. \cite{Shannon2018} and \cite{Macquart2018} found that the FRB population have a mean spectrum index $\alpha  = 1.8 \pm 0.3$ and $\alpha  = 1.6_{ - 0.3}^{ + 0.2}$, respectively. We adopt $\alpha  = 1.6$ in this paper.
We take ${\nu _1} = 1\,{\rm{GHz}}$ and ${\nu _2} = 8\,{\rm{GHz}}$, because an FRB at redshift $\sim4$ with characteristic frequency $\sim8\,{\rm{GHz}}$ can be detected in the observed band of Parkes and ASKAP. The results are presented in Figure 1.

As discussed by \cite{Keane2015}, the fluence completeness should be considered when one attempts to determine the population estimates for FRBs. The detection criteria is decided by the signal-to-noise ratio\citep{Caleb2016},
\begin{equation}
\textrm{S/N} = \frac{{\beta G\sqrt {B{N_{\rm{p}}}} }}{{{T_{{\rm{sys}}}}}}{S_\nu }\sqrt {{W_{{\rm{obs}}}}}  ~,
\label{eq:S/N}
\end{equation}
where $\beta$ is the digitisation factor, $G$ is the system gain in ${\rm{K}}{\kern 1pt} {\kern 1pt} {\rm{J}}{{\rm{y}}^{ - 1}}$, $B$ is the bandwidth in Hz, ${{N_{\rm{p}}}}$ is the number of polarizations, ${{T_{{\rm{sys}}}}}$ is the system temperature in ${\rm{K}}$, ${S_\nu }$ is the flux density of the signal in ${\rm{Jy}}$, and ${{W_{{\rm{obs}}}}}$ is the observed pulse width in seconds.
The signal is claimed as a reliable FRB detection when the $S/N$ reaches over some critical values,typically 8 to 10.
%The signal-to-noise ratio is usually adopted as $S/N>10$ for a reliable detection of an FRB.
Based on Eq~(\ref{eq:S/N}),then we have the detection threshold in fluence density,
\begin{equation}
{F_{\nu ,{\rm{th}}}} = \frac{{{T_{{\rm{sys}}}}}}{{\beta G\sqrt {B{N_{\rm{p}}}} }}(\textrm{S/N})\sqrt {{W_{{\rm{obs}}}}}   ~.
\label{eq:Futh}
\end{equation}
One can see that the detection threshold is depends on the pulse width ${{W_{{\rm{obs}}}}}$.
For the same $F_{\nu}$, a signal with larger ${{W_{{\rm{obs}}}}}$ is more difficult to be detected by the telescope.

\begin{figure}
\vskip-0.1in
\centerline{\includegraphics[angle=0,width=1.1\hsize]{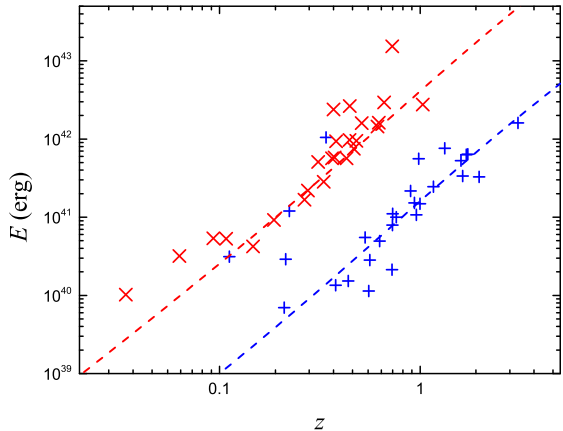}}
\vskip-0.1in
\caption{The energy--redshift distributions of 28 Parkes FRBs (blue data points) and 25 ASKAP FRBs (red data points).
The dash lines show the thresholds of ${F_{\nu {\rm{,th}}}} = 2\; {\kern 1pt} {\rm{Jy}}{\kern 1pt} {\rm{ms}}$ (blue) and ${F_{\nu {\rm{,th}}}} = 51\; {\kern 1pt} {\rm{Jy}}{\kern 1pt} {\rm{ms}}$ (red) for Parkes sample and ASKAP sample, respectively.}
\label{f1}
\end{figure}
According to Eq~(\ref{eq:Futh}), we only have fluence completeness for ${F_\nu } \gtrsim 2~{\kern 1pt} {\rm{Jy}}{\kern 1pt} {\rm{ms}}$ with a pulse width of $ \sim {\rm{32}}\,{\rm{ms}}$ which is the maximum to which FRB searches at Parkes are sensitive. We adopt a fluence threshold limit ${F_{\nu ,{\rm{th}}}} \simeq 2~{\kern 1pt} {\rm{Jy}}{\kern 1pt} {\rm{ms}}$, and pick a sub-sample containing 17 FRBs out of the Parkes sample containing a total of 28 FRBs.

For ASKAP, according to Eq~(\ref{eq:Futh}), \cite{Shannon2018} obtained a completeness threshold $\sim26$ Jyms with a pulse width matching the time resolution of 1.26 ms. However, in the current ASKAP sample, the pulse width has a typical value of several ms, with a maximum width of 5.4 ms for FRB 171019.
Therefore, we have a completeness threshold of ${\rm{51}}\,{\rm{Jyms}}$ by adopting ${W_{{\rm{obs}}}} \sim 5\,{\rm{ms}}$. Using the threshold  $\sim {\rm{51}}\,{\rm{Jyms}}$, we also select a sub-sample containing 21 FRBs out of the ASKAP sample containing a total of 25 FRBs.

In the $E - z$ plane, the truncated threshold ${F_{\nu ,{\rm{th}}}} \sim 2{\kern 1pt} {\rm{Jy}}{\kern 1pt} {\rm{ms}}$ for Parkes sample and ${F_{\nu ,{\rm{th}}}} \sim 51{\kern 1pt} {\rm{Jy}}{\kern 1pt} {\rm{ms}}$ for ASKAP sample are displayed by the red  and blue dash lines in Figure~\ref{f1}, respectively, and the corresponding energy threshold is calculated by Eq~(\ref{eq:Eiso}). One can see that the data points show a strong concentration toward the truncation line. In the next section, we estimate the energy function and cosmic formation rate of FRBs based on these two sub-samples, respectively.

\section{Energy function and formation rate}
For a survey of a telescope, the number of FRBs detectable above its threshold limit in the redshift range $({z_1},{z_2})$ and energy
range $({E_1},{E_2})$ can be expressed as
\begin{equation} \label{eq-N}
N = \frac{{\Omega T}}{{4\pi }}\int_{{z_1}}^{{z_2}} {dz} \int_{\max ({E_1},{E_{\min }})}^{{E_2}} {\frac{{\Psi (E,z)}}{{1 + z}}\frac{{dV}}{{dz}}\,}dE ~,
\end{equation}
Here $\Omega $ and $T$ are the field-of-view and the total observing time, respectively.
$dV$ is the co-moving volume element of the universe. $E_{\rm min}$  is the minimum observable energy at a redshift $z$.
${\Psi (E,z)}$ is the total energy function.

If ${\Psi (E,z)}$ is known, one can calculate the observed number of FRBs by a given telescope according to Equation \eqref{eq-N}.
If $E$ is independent of $z$, without loss of generality, ${\Psi (E,z)}$ can be expanded into a degenerate
(separate) form $\psi (E)\dot \rho (z)$, where $\psi (E)$ is the energy function and $\dot \rho (z)$ is the formation rate of FRBs.

However, as shown in Figure \ref{f1}, $E$ and $z$ appear to be highly correlated with each other both for Parkes and ASKAP sample.
Of course, this might be a result of the truncated effect and may not reflect the real physics.
Thus, the first step is to test the intrinsic correlation between $E$ and $z$. In order to achieve the goal, we use the Efron-Petrosian method \citep{Efron1992} which is a test of independence for truncated data. For the $i$th FRB in our sample, described by $(E_{i},\;{z_i})$, we can define the associated set as
\begin{equation} \label{eq-Ai}
{A_i} = \left\{ {j|E_{j} \ge E_{i},\;{z_j} \le {z_{\max ,i}}} \right\} ~,
\end{equation}
where $z_{{\rm max},i}$ is the maximum redshift at which the FRB with energy $E_{i}$
can be detected by the telescope above the detection threshold . The number of FRBs in this associated set $A_{i}$ is denoted as $N_{i}$.
We further define $A_{i}$'s largest un-truncated subset as
\begin{equation} \label{eq-Bi}
{B_i} = \left\{ {j \in {A_i}|\;{z_j} \le {z_i}} \right\} ~.
\end{equation}
The number of FRBs in $B_{i}$ is denoted as $R_{i}$.
Then we calculate the modified Kendall correlation coefficient through the statistic $\tau$:
\begin{equation} \label{eq-tau}
\tau  = \frac{{\sum\nolimits_i {({R_i} - {X_i})} }}{{\sqrt {\sum\nolimits_i {{V_i}} } }} ~,
\end{equation}
where ${X_i} = ({N_i} + 1)/2$ and ${V_i} = (N_i^2 - 1)/12$ are the expectation and variance for the uniformly scattering distribution, respectively.
Theoretically, a small $|\tau |$ value $(\leq1)$ implies energy evolution of FRBs is independent with redshift and $|\tau | \gg 1$ implies a strong dependence.
We find $|\tau | \simeq 0.4$ for Parkes sample which implies that $E$ is independent of $z$. And we find $|\tau | \simeq 1.1$ for
ASKAP sample, which means that $E$ is slightly correlated with $z$. However, we think that the independence between $E$ and $z$
is acceptable since the degree of correlation is very low.

Since $E$ is independent of $z$, we can derive the energy function ${\psi}(E)$ and formation rate $\dot \rho (z)$ of FRBs respectively by applying the $\rm C^{-}$ method proposed by \cite{Lynden-Bell1971}. This method has been widely used in the study of gamma-ray bursts (GRBs) \citep{Lloyd-Ronning2002, Yonetoku2004, Yonetoku2014, Petrosian2015, Yu2015, Deng2016, Zhang2018} and radio pulasrs \citep{2016Ap&SS.361..138D}. The robustness of this method has been confirmed by these works through Monte Carlo simulations.  \cite{Pescalli2016} further found that the $\rm C^{-}$ method can give more reliable results when applied to complete samples.

The cumulative energy distribution $\Phi  (>E)$ can be calculated point-by-point starting from the lowest observed energy \citep{Lynden-Bell1971},
\begin{equation} \label{eq-E0}
\Phi (>{E_{i}}) = \prod\limits_{j < i} {(1 - \frac{1}{{{N_j}}})}   ~,
\end{equation}
where $j<i$ implies that the $j$th FRB has a larger energy than the $i$th one.

\begin{figure}
\vskip-0.1in
\centerline{\includegraphics[angle=0,width=1.1\hsize]{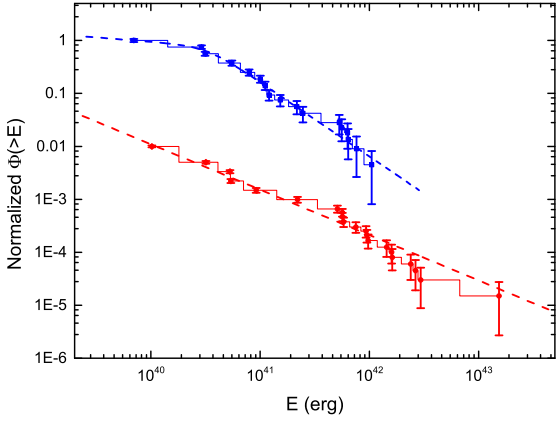}}
\vskip-0.1in
\caption{The normalized cumulative energy distributions of FRBs for Parkes sample (blue) and ASKAP sample (red), respectively. The uncertainties are the Poisson uncertainties of $N_{j}$. The dash line is the best fit to the data. For Parkes sample, the data can be best fitted by a broken power law function with $\chi^{2} / d o f=0.9$. For ASKAP sample, the best fitting function is a simple power law with $\chi^{2} / d o f=3.5$. One should note that the vertical axis of  ASKAP date have been multiplied by 0.01 just for the visual.}
\label{f2}
\end{figure}

The normalized cumulative energy distributions $\Phi(>E)$ for Parkes sample (blue step line) and ASKAP sample (red step line) are shown in Figure \ref{f2}. For Parkes sample, the best fitting function to the cumulative energy distributions is a broken power law. However, for ASKAP sample, the best fitting function is a simple power law. The differential energy function $\psi(E)$ can then be obtained by derivation of $\Phi(>E)$, yielding
\begin{equation}
\psi(E) \propto \left\{\begin{array}{ll}{E^{-1.3 \pm 0.1},} & {E<E_{b}} \\ {E^{-2.4 \pm 0.1},} & {E>E_{b}}\end{array}\right.
\end{equation}
for Parkes sample, where $E_{b}$ is the break energy and the fitting value is $E_{b}=(3.7 \pm 0.7) \times 10^{40}$ erg.
For ASKAP sample, the differential energy function is
\begin{equation} \label{eq-E}
\psi (E) \propto {E^{ - 1.9 \pm 0.1}}
\end{equation}

To derive the cosmic formation rate $\dot \rho (z)$ of FRBs, we define another associated set $C_i$ as,
\begin{equation}
{C_i} = \left\{ {j|{z_j} < {z_i},\;{E_{j}} \ge E_{i}^{{\rm{min}}}} \right\}   ~,
\label{eq-Ci}
\end{equation}
where $z_{i}$ is the redshift of $i$th FRB and $E_{0,i}^{\rm min}$ is the minimum observable energy at redshift $z_{i}$.
The number of FRBs in $C_{i}$ is denoted as $M_{i}$.
Similar to deriving the energy function, we can obtain the cumulative redshift distribution $\phi (<z)$ as \citep{Lynden-Bell1971},
\begin{equation} \label{eq-phiz}
\phi (<{z_i}) = \prod\limits_{j < i} {(1 + \frac{1}{{{M_j}}})}  ~.
\end{equation}
\begin{figure}
\vskip-0.1in
\centerline{\includegraphics[angle=0,width=1.1\hsize]{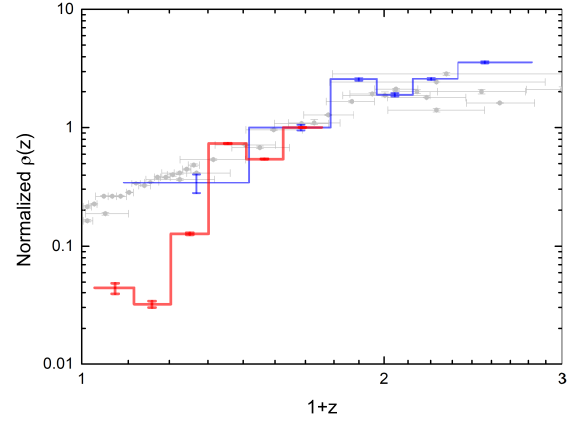}}
\vskip-0.1in
\caption{The cosmic formation rate $\dot \rho (z)$ of FRBs (step lines),
the error bars are obtained from the Poisson uncertainties of $M_{j}$. The blue step
line is derived by the Parkes sub-sample (17 FRBs). The red step line is derived by the
ASKAP sub-ample (21 FRBs). The cosmic star formation rate (gray points)
obtained from \cite{Hopkins2006} is also shown for comparison. All the data have been normalized to ${\rm{z}} \sim {\rm{0}}{\rm{.7}}$.}
\label{f3}
\end{figure}
We can derive the FRB cosmic formation rate $\dot \rho (z)$ with the following formula,
\begin{equation} \label{eq-rhoz}
\dot \rho (z) \propto (1 + z)\frac{{d\phi (<z)}}{{dz}}{\left( {\frac{{dV}}{{dz}}} \right)^{ - 1}}  ~,
\end{equation}
where the factor $(1 + z)$ comes from the cosmological time dilation.

Figure \ref{f3} gives the normalized cosmic formation rate derived from Parkes sample (blue step line) and ASKAP sample (red step line), respectively, comparing with the normalized cosmic star formation rate (gray points). One can see that the formation rate derived from Parkes sample is roughly consistent with the star formation rate (SFR) up to $z\sim1.7$. However,  the formation rate derived from ASKAP sample evolves much faster than the SFR up to $z\sim0.7$, with $\dot \rho (z) \propto {(1 + z)^{6.9 \pm 1.9}}$.

Moreover, one can obtain the local formation rate $\dot \rho (0) \simeq 3.2\pm 0.3\; \times {10^4}\;{\rm{Gp}}{{\rm{c}}^{ - 3}}{\rm{y}}{{\rm{r}}^{ - 1}}$ for Parkes sample and $\dot \rho (0) \simeq 4.6\pm 0.8\; \times {10^3}\;{\rm{Gp}}{{\rm{c}}^{ - 3}}{\rm{y}}{{\rm{r}}^{ - 1}}$ for ASKAP sample, respectively, by assuming no beaming effect. Here we have adopted the total exposure $\Omega T \simeq (700 + 1115 + 907) \times 0.55 \times 0.9{~\deg ^2}{\rm{h}}{\mkern 1mu}  \simeq {\rm{1347}}.{\rm{4}}{~\deg ^2}{\rm{h}}$ based on the SUPERB survey for Parkes sample \citep{Bhandari2018}, where we have assumed a 90\% survey efficiency. And we have adopted the total exposure $\Omega T \simeq 5.1 \times {10^5}{~\deg ^2}{\rm{h}}$ for ASKAP sample \citep{Shannon2018}.

\section{CONCLUSION and DISCUSSION}
In this work, we use the Lynden-Bell $\rm C^{-}$ method to study the released energy function and cosmic formation rate of FRBs without model assumptions.
This is the first time for applying this method to FRBs. Firstly, we find that the energy of FRBs is independent of the redshift.
We derive the differential energy function, it is a broken power law for Parkes sample, however it is a simple power law for the ASKAP sample. Moreover, we also derive the cosmic formation rate of FRBs.
It is a surprise that the formation rate of FRBs derived from Parkes sample is roughly consistent with the SFR up to $z\sim1.7$,
with a local rate of $\dot \rho (0) \simeq (3.2 \pm 0.3)\; \times {10^4}\;{\rm{Gp}}{{\rm{c}}^{ - 3}}{\rm{y}}{{\rm{r}}^{ - 1}}$. However, we find that
the formation rate of FRBs derived from ASKAP sample evolves much faster than the SFR up to $z\sim0.7$ namely $\dot \rho (z) \propto {(1 + z)^{6.9 \pm 1.9}}$, with a local rate of $\dot \rho (0) \simeq (4.6 \pm 0.8)\; \times {10^3}\;{\rm{Gp}}{{\rm{c}}^{ - 3}}{\rm{y}}{{\rm{r}}^{ - 1}}$.
Note that the local rate derived from the Parkes sample is larger than that from the ASKAP sample by a factor of $\sim7$.  This is understandable because Parkes is much sensitivity than ASKAP.

We note that \cite{Locatelli2018} also found that the formation rate of FRBs is consistent with the cosmic SFR for Parkes sample, but it evolves faster than the SFR for ASKAP sample. They could not tell whether such fast evolution (ASKAP sample) is due to an intrinsic density evolution or to a luminosity (or energy) evolution. However, in our work, we show that the energy evolution of FRBs is likely independent of the redshift\footnote{One should note that the redshifts are inferred by the DM. Therefore, this conclusion is biased by the accuracy of the inferred redshifts.}. Therefore, this fast evolution may not be not due to a luminosity (or energy) evolution.

It is odd that we obtain different results, namly the energy function and cosmic formation rate, from Parkes sample and ASKAP sample by using the same method.
One thing must be noted is that the fluences measured by Parkes are actually lower limits due to the unknown position of the FRBs in the beam pattern \citep{Macquart2018a}. This will affect the energy estimation for Parkes sample, and this effect is difficult to estimate. However, the fluences measured by AKSAP are much more reliable than Parkes since they use a phased array feed. But unfortunately, the ASKAP sample probes a much smaller volume than Parkes sample. Also, the small sizes of the current sample of FRBs limit the robustness of the results in this paper. More FRBs samples are needed to further study this subject in the future work. For instance, it is expected that CHIME can detect 2-42 FRBs per day \citep{CHIME/FRB Collaboration2018}. We believe that, not too long, a larger sample from the CHIME telescope could yield interesting results on the FRB population as a whole.

Anyway, we derived a local rate $\dot \rho (0) \sim {10^4}\;{\rm{Gp}}{{\rm{c}}^{ - 3}}{\rm{y}}{{\rm{r}}^{ - 1}}$ for non-repeating FRBs. It allows us to
compare this rate to various progenitors model.  Furthermore, the formation rate is likely to evolve fast at low redshift. In any case, FRB models should explain the formation rate both at local and its evolution.

\section{ACKNOWLEDGMENTS}

We thank S.-B. Zhang for helpful discussion. This work is partially supported by the National Natural Science Foundation of China
(grant Nos. 11603076, 11673068, 11725314, and U1831122), the Youth Innovation Promotion Association (grant No. 2017366), the Key Research Program of Frontier Sciences (QYZDB-SSW-SYS005), the Strategic Priority Research Program ``Multi-waveband gravitational wave Universe'' (grant No. XDB23000000) of the Chinese Academy of Sciences, and the ``333 Project'' and the Natural Science Foundation (Grant No. BK20161096) of Jiangsu Province.

\begin{longtable}{lccccccc}
	\caption{The observational properties and the estimated redshifts ($z$) and isotropic energies ($E$) of FRBs.}\\
	\hline
	\hline
	FRB Name & $\rm DM_{obs}$      & $\rm DM_{MW}$       &   ${z}$   &  $F_{\nu}$  &  ${{\tau_{{\rm{obs}}}}}$    &  $E$     \\
	& (pc $\rm cm^{-3}$)  & (pc $\rm cm^{-3}$)  &         &  (Jyms)          &  (ms)                    &  ($10^{40}$ erg)\\
	
	%\begin{deluxetable}{ccccccc}
		%\tablewidth{350pt} %\rotate
	%	\tabletypesize{\footnotesize}
		%\tablecaption{Results of smoothly broken power-law function fits to the GRB cumulative luminosity distributions in given redshift ranges for GRBs in our sample.}
	%	\tablehead{ \colhead{ FRB Name}&
		%	\colhead{$\rm DM_{obs$}& \colhead{$\rm DM_{MW}$}& \colhead{${z}$}& \colhead{$F_{\nu}$}& \colhead{${{W_{{\rm{obs}}}}}$}& \colhead{$E$}
		%	}
		%\startdata
		%\hline
	\hline
	\hline
	Parkes FRBs &      &      &     &   &      &       \\
	\hline
	FRB010125	&$	790	$&	110	&$	0.73	$&	2.82	&	9.4	&$	11.08	$&	\\
	FRB010621	&$	745	$&	523	&$	0.21	$&	2.87	&	7	&$	2.91	$&	\\
	FRB010724	&$	375	$&	44.58	&$	0.34	$&	150	&	5	&$	105.04	$&	\\
	FRB090625	&$	899.55	$&	31.69	&$	0.94	$&	2.1888	&	1.92	&$	15.28	$&	\\
	FRB110214	&$	168.9	$&	31.1	&$	0.11	$&	51.3	&	1.9	&$	3.13	$&	\\
	FRB110220	&$	944.38	$&	34.77	&$	0.98	$&	7.28	&	5.6	&$	55.87	$&	\\
	FRB110626	&$	72	$&	47.46	&$	0.72	$&	0.56	&	1.4	&$	2.13	$&	\\
	FRB110703	&$	1103.6	$&	32.33	&$	1.17	$&	2.15	&	4.3	&$	24.63	$&	\\
	FRB120127	&$	553.3	$&	31.82	&$	0.55	$&	0.55	&	1.1	&$	1.14	$&	\\
	FRB121002	&$	1629.18	$&	74.27	&$	1.73	$&	2.3392	&	5.44	&$	64.05	$&	\\
	FRB130626	&$	952.4	$&	66.87	&$	0.96	$&	1.4652	&	1.98	&$	10.73	$&	\\
	FRB130628	&$	469.88	$&	52.58	&$	0.44	$&	1.2224	&	0.64	&$	1.52	$&	\\
	FRB130729	&$	861	$&	31	&$	0.9	$&	3.4342	&	15.61	&$	21.72	$&	\\
	FRB131104	&$	779	$&	71.1	&$	0.76	$&	2.3296	&	2.08	&$	10.03	$&	\\
	FRB140514	&$	562.7	$&	34.9	&$	0.56	$&	1.3188	&	2.8	&$	2.84	$&	\\
	FRB150215	&$	1105.6	$&	427.2	&$	0.73	$&	2.016	&	2.88	&$	7.92	$&	\\
	FRB150418	&$	776.2$&	188.5	&$	0.63	$&	1.76	&	0.8	&$	4.95	$&	\\
	FRB150610	&$	1593.9	$&	122	&$	1.63	$&	1.4	&	2	&$	33.62	$&	\\
	FRB150807	&$	266.5	$&	36.9	&$	0.22	$&	44.8	&	0.35	&$	12.04	$&	\\
	FRB151206	&$	1909.8	$&	160	&$	1.97	$&	0.9	&	3	&$	32.75	$&	\\
	FRB151230	&$	960.4	$&	38	&$	1	$&	1.848	&	4.4	&$	14.85	$&	\\
	FRB160102	&$	2596.1	$&	13	&$	3.08	$&	1.7	&	3.4	&$	160.82	$&	\\
	FRB171209	&$	1458	$&	13	&$	1.6	$&	2.3	&	2.5	&$	53.01	$&	\\
	FRB180301	&$	520	$&	155	&$	0.38	$&	1.5	&	3	&$	1.35	$&	\\
	FRB180309	&$	263.47	$&	44.69	&$	0.21	$&	11.9808	&	0.576	&$	0.7	$&	\\
	FRB180311	&$	1575.6	$&	45.2	&$	1.7	$&	2.4	&	12	&$	63.23	$&	\\
	FRB180714	&$	1469.873	$&	257	&$	1.33	$&	5	&	1	&$	76.37	$&	\\
	FRB180923	&$	548	$&	46.6	&$	0.53	$&	2.9	&	20	&$	5.51	$&	\\
	\hline
	ASKAP FRBs &      &      &     &   &      &       \\
	\hline
	FRB170107	&$	609.5	$&	35	&$	0.61	$&	58	&	2.4	&$	144.54	$&	\\
	FRB170416	&$	523.2	$&	40	&$	0.51	$&	97	&	5	&$	158.49	$&	\\
	FRB170428	&$	991.7	$&	40	&$	1.03	$&	34	&	4.4	&$	275.42	$&	\\
	FRB170707	&$	235.2	$&	36	&$	0.19	$&	52	&	3.5	&$	9.12	$&	\\
	FRB170712	&$	312.79	$&	38	&$	0.28	$&	53	&	1.4	&$	21.88	$&	\\
	FRB170906	&$	390.3	$&	39	&$	0.36	$&	74	&	2.5	&$	56.23	$&	\\
	FRB171003	&$	463.2	$&	40	&$	0.44	$&	81	&	2	&$	97.72	$&	\\
	FRB171004	&$	304	    $&	38	&$	0.27	$&	44	&	2	&$	16.6	$&	\\
	FRB171019	&$	460.8	$&	37	&$	0.45	$&	219	&	5.4	&$	263.03	$&	\\
	FRB171020	&$	114.1	$&	38	&$	0.03	$&	200	&	3.2	&$	1.02	$&	\\
	FRB171116	&$	618.5	$&	36	&$	0.62	$&	63	&	3.2	&$	162.18	$&	\\
	FRB171213	&$	158.6	$&	36	&$	0.09	$&	133	&	1.5	&$	5.37	$&	\\
	FRB171216	&$	203.1	$&	37	&$	0.15	$&	40	&	1.9	&$	4.27	$&	\\
	FRB180110	&$	715.7	$&	38	&$	0.73	$&	420	&	3.2	&$	1548.82	$&	\\
	FRB180119	&$	402.7	$&	36	&$	0.38	$&	110	&	2.7	&$	93.33	$&	\\
	FRB180128.0	&$	441.4	$&	32	&$	0.43	$&	51	&	2.9	&$	56.23	$&	\\
	FRB180128.2	&$	495.9	$&	40	&$	0.48	$&	66	&	2.3	&$	95.5	$&	\\
	FRB180130	&$	343.5	$&	39	&$	0.31	$&	95	&	4.1	&$	51.29	$&	\\
	FRB180131	&$	657.7	$&	40	&$	0.66	$&	100	&	4.5	&$	295.12	$&	\\
	FRB180212	&$	167.5	$&	33	&$	0.11	$&	96	&	1.81&$	5.25	$&	\\
	FRB180315	&$	479   	$&	36	&$	0.47	$&	56	&	2.4	&$	75.86	$&	\\
	FRB180324	&$	431	    $&	70	&$	0.37	$&	71	&	4.3	&$	57.54	$&	\\
	FRB180430	&$	264.1	$&	165.44	&$0.06  $&	177	&	1.2	&$	3.16	$&	\\
	FRB180515	&$	355.2	$&	33  &$	0.3	    $&	46	&	1.9	&$	28.18	$&	\\
	FRB180525	&$	388.1	$&	31  &$	0.4	    $&	300	&	3.8	&$	239.88	$&	\\
	\hline
	\hline
	
	\end{longtable}

\end{document}